\documentclass[11pt]{article}

\usepackage[T1]{fontenc}
\usepackage[utf8]{inputenc}
\usepackage{lmodern}
\usepackage[a4paper,margin=2.45cm]{geometry}
\usepackage{microtype}
\usepackage{amsmath,amssymb,mathtools,bm}
\usepackage{booktabs,array}
\usepackage{enumitem}
\usepackage[numbers,sort&compress]{natbib}
\usepackage{xcolor}
\usepackage{hyperref}
\usepackage[T1]{fontenc}
\usepackage[utf8]{inputenc}   % Only for pdfLaTeX
\usepackage{lmodern}
\usepackage{cmap}

\hypersetup{
  colorlinks=true,
  linkcolor=blue!45!black,
  citecolor=blue!45!black,
  urlcolor=blue!55!black,
  pdftitle={Horizon-Boundary Localization of the Hawking Temperature in a Weak AdS/CFT Extension},
  pdfauthor={Wen-Xiang Chen}
}

\setlist{nosep,leftmargin=2.1em}
\numberwithin{equation}{section}

\newtheorem{proposition}{Proposition}

\newcommand{\dd}{\mathrm{d}}
\newcommand{\ii}{\mathrm{i}}
\newcommand{\e}{\mathrm{e}}
\newcommand{\Res}{\operatorname*{Res}}

\newcommand{\Order}{\mathcal{O}}
\newcommand{\AdS}{\mathrm{AdS}}
\newcommand{\CFT}{\mathrm{CFT}}
\newcommand{\TBH}{T_{\mathrm H}}
\newcommand{\betaH}{\beta_{\mathrm H}}

\title{\textbf{Horizon-Boundary Localization of the Hawking Temperature\\
in a Weak AdS/CFT Extension}\\[0.35em]}
\author{Wen-Xiang Chen$^{1}$ and M. Zeeshan Gul$^{2,3,4}$\\
\small $^{1}$School of Electronic Information, Guangzhou City University of Technology,\\$^{1}$wxchen4277@qq.com\\
\small $^2$ Department of Mathematics and Statistics, The
University of Lahore, \\
\small 1-KM Defence Road Lahore-54000, Pakistan\\
\small $^{3}$ Research Center of Astrophysics and Cosmology, Khazar
University, \\
\small Baku, AZ1096, 41 Mehseti Street, Azerbaijan\\
\small $^{4}$ Jadara Research Center, Jadara University, Irbid 21110, Jordan\\$^2$mzeeshangul.math@gmail.com}

\date{}

\begin{document}
\maketitle

\begin{abstract}
We formulate and prove a conditional boundary-localization statement for the Hawking
temperature.  For a static, Euclideanizable black hole whose lapse functions have a
non-degenerate common zero at $r=r_h$, the near-horizon geometry is a smooth disk only
when the Euclidean period is
$\beta_{\mathrm H}=4\pi/\sqrt{A'(r_h)B'(r_h)}$.  We derive this result step by step in
three mutually consistent languages: removal of the conical defect, the horizon boundary
term in the two-dimensional Gauss--Bonnet theorem, and the residue of the meromorphic
one-form
$\omega_h=\dd z/\sqrt{A(z)B(z)}$.  In particular,
\[
  \beta_{\mathrm H}=4\pi\left|\Res_{z=r_h}\omega_h\right|,
  \qquad
  T_{\mathrm H}=\frac{1}{4\pi
  \left|\Res_{z=r_h}\omega_h\right|}.
\]
The complex contour probes the same local horizon data as the limiting boundary of an
excised Euclidean disk; these are distinct curves.  Hence, with time normalization fixed,
the temperature is fixed entirely by the first non-vanishing horizon data.  In an asymptotically
AdS saddle admitting the usual thermal holographic dictionary, the same Euclidean circle
reaches the conformal boundary.  The boundary CFT therefore inherits the horizon-selected KMS period, up to the chosen normalization of
boundary time.  This is a weak extension of the AdS/CFT organizing principle restricted
to the temperature sector; it neither identifies the horizon with the AdS conformal
boundary nor establishes a full duality.  For a dynamical spherically symmetric metric,
the outgoing Hamilton--Jacobi momentum at finite nonzero Kodama energy has a simple pole
at $C(v,r)=0$, the non-degenerate future outer trapping/apparent horizon.  A separated event horizon instead obeys
$C(v,r_E)=2\e^{-\Psi}\dot r_E$ and is generically not a pole.  Vaidya and charged-Vaidya
examples thus provide a boundary-selector test: at leading adiabatic order the local
temperature is attached to the trapping-horizon boundary, while all relevant horizons
coincide in the stationary limit of these examples.  For a general static limit,
the Kodama and Killing temperatures are related by the horizon redshift factor
$T_{\mathrm{Kill}}=\e^{\Psi_H}T_{\mathrm{Kod}}$.  Schwarzschild--AdS and Reissner--Nordstr\"om--AdS
examples, numerical contour integrals, the flat and neutral limits, and an extremal
counterexample are supplied as checks.  The result concerns the localization of the
temperature parameter, not a replacement for Hawking's quantum-field-theoretic particle
creation mechanism.
\end{abstract}

\section{Introduction}

Hawking's quantum-field-theoretic calculation established that a stationary black hole
emits a thermal spectrum with temperature $\TBH=\kappa/(2\pi)$, where $\kappa$ is the
surface gravity in a specified normalization of time \citep{Hawking1975}.  The path-integral
and Euclidean approaches identify the same temperature with the period required to remove
a conical singularity at the Euclidean horizon \citep{HartleHawking1976,GibbonsHawking1977}.
These results distinguish two questions that are sometimes conflated: the quantum mechanism
that produces particles and the local geometric data that fix the temperature appearing in
the thermal factor.  This paper addresses the second question.

The AdS/CFT correspondence relates gravitational saddles in asymptotically anti-de Sitter
spacetimes to thermal states and generating functionals of a conformal field theory on the
asymptotic boundary \citep{Maldacena1998,Gubser1998,Witten1998Holography}.  In the
finite-temperature correspondence, the Euclidean time circle of an AdS black-hole saddle
is also the thermal circle of the boundary theory \citep{HawkingPage1983,Witten1998Thermal}.
The same period occurs at two geometrically different locations: it is constrained by
regularity at the horizon and is read as a KMS period at the conformal boundary.  It is
therefore essential not to call both surfaces simply ``the boundary.''

A second line of work expresses black-hole temperature through the Euler characteristic of
the dimensionally reduced Euclidean geometry.  The Robson--Villari--Biancalana (RVB)
method gives a topological temperature formula for two-dimensional black holes
\citep{Robson2019}.  A residue-based development reorganizes the near-horizon contribution
in terms of complex singularity data \citep{Chen2025,Chen2026Dynamic}.  Related work
connects horizon residues with thermodynamic geometry and quasinormal-mode monodromy
\citep{Chen2026Weak}.  Here we identify the differential-geometric and complex-analytic
descriptions only through their common Hawking-temperature invariant and restrict the
proposed weak AdS/CFT extension to that temperature sector.  This leaves a sharper question:

\begin{quote}
\emph{Can one show, rather than only infer from equal final formulas, that the temperature
is localized on a definite horizon boundary and that the AdS boundary merely inherits the
corresponding period?}
\end{quote}

The distinction becomes operational in a dynamical spacetime.  Event and apparent horizons
need not coincide.  In the tunnelling description of evaporating Vaidya-type black holes,
they can form the two edges of a quantum ergoregion, and the apparent horizon has been
identified as the surface from which the radiation becomes detectable
\citep{ZhangZhao2010}.  Independently, trapping-horizon mechanics and covariant
Hamilton--Jacobi tunnelling attach a local temperature to a future outer trapping horizon
\citep{Hayward1994,Hayward2009,DiCriscienzo2010}.  These results supply a decisive test of
the proposed residue dictionary: the relevant boundary should be the horizon at which the
radial action actually develops its pole.

The contributions of the present paper are fourfold.
\begin{enumerate}
  \item We prove a static boundary-localization proposition for a general diagonal
  two-dimensional sector with distinct lapse functions $A(r)$ and $B(r)$.
  \item We show explicitly that conical regularity, the Gauss--Bonnet horizon term, and the
  residue of $\dd z/\sqrt{A(z)B(z)}$ give the same period without confusing a dimensionless
  winding number with a dimensionful temperature scale.
  \item We formulate the weak holographic map as a map of Euclidean periods.  The horizon
  regulator contour, the complex contour, and the AdS conformal boundary are kept distinct.
  \item We derive a dynamical boundary-selector criterion.  For generalized Vaidya
  coordinates, the Hamilton--Jacobi pole lies at the trapping/apparent horizon and not, in
  general, at a separated event horizon.
\end{enumerate}

Throughout, $G=\hbar=c=k_{\mathrm B}=1$.  The word ``proof'' refers to the propositions
under their stated geometric, analytic, semiclassical, and adiabatic assumptions.  It does
not refer to a proof of a complete AdS/CFT duality or to a new derivation of the full quantum
state of Hawking radiation.

\section{Framework, assumptions, and the precise claim}
\label{sec:framework}

\subsection{Static exterior and time normalization}

Consider the static two-dimensional sector of a spherically symmetric black-hole exterior,
\begin{equation}
  \dd s^2_{(2)}=-A(r)\,\dd t^2+\frac{\dd r^2}{B(r)}.
  \label{eq:static2d}
\end{equation}
The angular part is a spectator for the local temperature.  We assume that the outer horizon
$r=r_h$ is a common, non-degenerate zero,
\begin{equation}
  A(r_h)=B(r_h)=0,
  \qquad
  a_1\equiv A'(r_h)>0,
  \qquad
  b_1\equiv B'(r_h)>0.
  \label{eq:simplezeros}
\end{equation}
The signs shown are appropriate immediately outside an outer black-hole horizon.  Absolute
values will restore the corresponding formula for the opposite orientation.

The normalization of $t$ is physical.  Under a constant rescaling
\begin{equation}
  t_b=\alpha t,
  \label{eq:timerescale}
\end{equation}
with $\alpha>0$, the coefficient of $\dd t_b^2$ is $A_b=A/\alpha^2$.
Consequently, temperatures paired with this static Killing-time energy transform as
$T_b=T/\alpha$.  The invariant Kodama temperature used in the dynamical section has a
different energy convention, as clarified below.  In asymptotically flat space, $\alpha$ is usually
fixed by unit norm of the Killing vector at infinity.  In AdS/CFT, it is fixed by the chosen
representative of the boundary conformal metric.

\subsection{Assumption ledger}

The result uses the following assumptions; none is hidden in the word ``boundary.''

\begin{table}[ht]
\centering
\caption{Assumptions and conventions for the boundary-localization statement.}
\label{tab:assumptions}
\small
\begin{tabular}{p{0.19\linewidth}p{0.73\linewidth}}
\toprule
Item & Assumption or convention \\
\midrule
Geometry & The relevant $(t,r)$ sector is regular outside the horizon and can be
Euclideanized locally. \\
Horizon & $A$ and $B$ have a common simple zero as in Eq.~\eqref{eq:simplezeros}. \\
Analyticity & $H(z)=\sqrt{A(z)B(z)}$ admits a single-valued analytic branch in a punctured
neighborhood of $z=r_h$ and has a simple zero there. \\
Contour & The contour surrounds only the selected horizon zero and crosses no pole, zero,
or branch cut during deformation. \\
Coordinates & Radial complex-coordinate changes are biholomorphic at the pole.  Singular
maps are excluded. \\
Time & The normalization of $t$ is fixed before comparing bulk, residue, and boundary-CFT
temperatures. \\
Quantum regime & The geometry is treated semiclassically.  Backreaction affects the result
only through the background functions unless stated otherwise. \\
Dynamics & Spherical symmetry, a simple future outer trapping horizon, and
$|\dot\kappa_H|/\kappa_H^2\ll1$ in a specified exterior-time normalization are imposed.
The leading-pole calculation freezes the background locally; $C(v,z)$ is analytic near
the selected zero and the horizon Kodama energy is finite and nonzero. \\
Claim strength & The correspondence is restricted to the thermal period and directly
dependent horizon data.  No Hilbert-space or operator-level duality is asserted. \\
\bottomrule
\end{tabular}
\end{table}

\subsection{Three different boundaries}

We shall use three related but inequivalent objects.
\begin{enumerate}
  \item $\mathcal C_\epsilon$ is the real one-dimensional inner boundary created by excising
  a disk of proper radius $\epsilon$ around the Euclidean horizon.
  \item $\Gamma_\epsilon$ is a complex contour surrounding the zero $z=r_h$.  It carries the
  residue data of the meromorphic one-form.
  \item $\partial\AdS$ is the asymptotic conformal boundary on which the CFT lives.
\end{enumerate}
The first two shrink onto the same local horizon germ.  The third is far away.  They share a
thermal period only because a single Euclidean time coordinate extends through the regular
bulk saddle.

\begin{proposition}[Static horizon-boundary localization]
\label{prop:static}
Under the assumptions in Table~\ref{tab:assumptions}, define
\begin{equation}
  H(z)=\sqrt{A(z)B(z)},
  \qquad
  \omega_h=\frac{\dd z}{H(z)}.
  \label{eq:H_and_omega}
\end{equation}
Then the regular Euclidean period, surface gravity, residue, and Hawking temperature satisfy
\begin{equation}
  \kappa_h=\frac{1}{2}\sqrt{a_1b_1},
  \qquad
  \betaH=\frac{2\pi}{\kappa_h}
        =4\pi\left|\Res_{z=r_h}\omega_h\right|,
  \qquad
  \TBH=\frac{\kappa_h}{2\pi}
     =\frac{1}{4\pi\left|\Res_{z=r_h}\omega_h\right|}.
  \label{eq:master_static}
\end{equation}
All quantities on the right-hand side depend only on the first non-vanishing horizon
coefficients and the normalization of time.
\end{proposition}

The next three sections provide independent, step-by-step derivations of
Proposition~\ref{prop:static}.

\section{Euclidean regularity: derivation from the horizon disk}
\label{sec:euclidean}

Perform the Wick rotation $t=-\ii\tau$.  Equation~\eqref{eq:static2d} becomes
\begin{equation}
  \dd s^2_{(2),E}=A(r)\,\dd\tau^2+\frac{\dd r^2}{B(r)}.
  \label{eq:euclidean_metric}
\end{equation}
Set $x=r-r_h$.  Taylor expansion at the simple horizon gives
\begin{align}
  A(r)&=a_1x+a_2x^2+\Order(x^3),
  \label{eq:Aexpansion}\\
  B(r)&=b_1x+b_2x^2+\Order(x^3).
  \label{eq:Bexpansion}
\end{align}
The coordinate $x$ is not the proper radial distance because the coefficient of
$\dd x^2$ diverges.  Define instead a radial coordinate that agrees with proper distance
at leading order,
\begin{equation}
  \rho=2\sqrt{\frac{x}{b_1}},
  \qquad
  x=\frac{b_1}{4}\rho^2,
  \qquad
  \dd x=\frac{b_1}{2}\rho\,\dd\rho.
  \label{eq:rho_definition}
\end{equation}
Substitution into the radial term yields
\begin{align}
  \frac{\dd r^2}{B(r)}
  &=\frac{(b_1^2\rho^2/4)\dd\rho^2}
  {b_1(b_1\rho^2/4)+\Order(\rho^4)}
  \nonumber\\
  &=\left[1+\Order(\rho^2)\right]\dd\rho^2.
  \label{eq:radial_regular}
\end{align}
The Euclidean-time term becomes
\begin{align}
  A(r)\dd\tau^2
  &=\left[\frac{a_1b_1}{4}\rho^2+\Order(\rho^4)\right]\dd\tau^2
  \nonumber\\
  &=\left[\kappa_h^2\rho^2+\Order(\rho^4)\right]\dd\tau^2,
  \label{eq:angular_regular}
\end{align}
where
\begin{equation}
  \kappa_h\equiv\frac{1}{2}\sqrt{a_1b_1}.
  \label{eq:kappa_ab}
\end{equation}
Consequently,
\begin{equation}
  \dd s^2_{(2),E}=\dd\rho^2+\rho^2\dd\theta^2+\Order(\rho^2)\dd\rho^2
  +\Order(\rho^4)\dd\tau^2,
  \qquad
  \theta\equiv\kappa_h\tau.
  \label{eq:polar_metric}
\end{equation}

The leading metric is the flat plane in polar coordinates.  The point $\rho=0$ is regular
only if the angular coordinate has period $2\pi$,
\begin{equation}
  \theta\sim\theta+2\pi.
  \label{eq:theta_period}
\end{equation}
If $\tau\sim\tau+\betaH$, Eq.~\eqref{eq:theta_period} becomes
\begin{equation}
  \kappa_h\betaH=2\pi.
  \label{eq:regularity_condition}
\end{equation}
Solving first for the period and then taking its inverse gives
\begin{equation}
  \betaH=\frac{2\pi}{\kappa_h}
        =\frac{4\pi}{\sqrt{a_1b_1}},
  \qquad
  \TBH=\betaH^{-1}=\frac{\sqrt{a_1b_1}}{4\pi}.
  \label{eq:euclidean_temperature}
\end{equation}

This derivation uses no data at a finite distance from the horizon.  For every sufficiently
small $\epsilon$, the circle $\mathcal C_\epsilon:\rho=\epsilon$ has circumference
\begin{equation}
  \ell(\mathcal C_\epsilon)
  =\int_0^{\betaH}\left[\kappa_h\epsilon+\Order(\epsilon^3)\right]\dd\tau
  =\kappa_h\betaH\epsilon+\Order(\epsilon^3).
  \label{eq:circumference}
\end{equation}
Regularity requires
$\lim_{\epsilon\to0}\ell(\mathcal C_\epsilon)/\epsilon=2\pi$.
Thus the temperature is a boundary condition on a shrinking horizon circle.

\section{Gauss--Bonnet and the RVB horizon term}
\label{sec:GB}

The preceding result can be written as a boundary statement in differential geometry.  For
a compact oriented two-manifold $\mathcal M_2$ with smooth boundary, the Gauss--Bonnet
theorem is \citep{Chern1944}
\begin{equation}
  \chi(\mathcal M_2)
  =\frac{1}{4\pi}\int_{\mathcal M_2}R\sqrt{g}\,\dd^2x
  +\frac{1}{2\pi}\int_{\partial\mathcal M_2}k_g\,\dd s,
  \label{eq:GB}
\end{equation}
up to the standard corner terms, which are absent for the circular regulator used here.

We now calculate the horizon boundary contribution without skipping the normalization
steps.  At fixed $r$, a unit tangent to the Euclidean-time circle and an outward unit normal
are
\begin{equation}
  u^a=\left(A^{-1/2},0\right),
  \qquad
  n^a=\left(0,B^{1/2}\right).
  \label{eq:unit_vectors}
\end{equation}
The only Christoffel symbol needed for the radial acceleration is
\begin{equation}
  \Gamma^r_{\tau\tau}
  =-\frac{1}{2}g^{rr}\partial_r g_{\tau\tau}
  =-\frac{1}{2}B A'.
  \label{eq:Gamma_rtt}
\end{equation}
The acceleration of the unit tangent is therefore
\begin{equation}
  a^r=u^b\nabla_bu^r
  =\Gamma^r_{\tau\tau}(u^\tau)^2
  =-\frac{BA'}{2A}.
  \label{eq:acceleration}
\end{equation}
Since $n_r=g_{rr}n^r=B^{-1/2}$, the geodesic curvature compatible with the positive
boundary term in Eq.~\eqref{eq:GB} is minus the acceleration projected onto the outward
normal,
\begin{equation}
  k_g=-n_a a^a=\frac{A'\sqrt{B}}{2A}.
  \label{eq:kg}
\end{equation}
The line element on the circle is $\dd s=\sqrt{A}\,\dd\tau$, so
\begin{equation}
  k_g\,\dd s=\frac{A'\sqrt{B}}{2\sqrt{A}}\,\dd\tau.
  \label{eq:kgds}
\end{equation}
Using Eqs.~\eqref{eq:Aexpansion} and \eqref{eq:Bexpansion},
\begin{align}
  \lim_{r\to r_h^+}\frac{A'\sqrt{B}}{2\sqrt{A}}
  &=\lim_{x\to0^+}\frac{[a_1+\Order(x)]
  \sqrt{b_1x+\Order(x^2)}}{2\sqrt{a_1x+\Order(x^2)}}
  \nonumber\\
  &=\frac{1}{2}\sqrt{a_1b_1}
  =\kappa_h.
  \label{eq:kglimit}
\end{align}
The sign in Eq.~\eqref{eq:kgds} reverses when the same circle is treated as the inner
boundary of the exterior region.  Its orientation-independent magnitude is therefore
\begin{equation}
  \left|\int_{\mathcal C_\epsilon}k_g\,\dd s\right|
  \xrightarrow[\epsilon\to0]{}\kappa_h\betaH.
  \label{eq:horizon_boundary_term}
\end{equation}

For a small disk that can be smoothly filled at $\rho=0$, $\chi=1$, while the regular bulk
curvature integral is $\Order(\epsilon^2)$.  Here the circle is the outer boundary of the
filled disk, so its signed contribution is positive.  An exterior annulus instead has
$\chi=0$ and requires both of its boundary contributions.  Taking the small-disk limit
in Eq.~\eqref{eq:GB} gives
\begin{equation}
  1=\frac{1}{2\pi}\kappa_h\betaH,
  \label{eq:GB_period}
\end{equation}
which is exactly Eq.~\eqref{eq:regularity_condition}.  If the period is arbitrary, the tip is
a cone with deficit
\begin{equation}
  \delta=2\pi-\kappa_h\beta,
  \label{eq:deficit}
\end{equation}
and a distributional Gaussian-curvature integral $\delta$, or equivalently a scalar-curvature
integral $2\delta$, restores the Gauss--Bonnet identity.  Hence topology alone does not select the physical period: the
additional physical condition is that the Euclidean saddle be regular, $\delta=0$.  This
qualification prevents a circular or over-strong use of the RVB formula.

For the Schwarzschild gauge $A=B=F$, the Euclidean metric is
\begin{equation}
  \dd s^2_{(2),E}=F(r)\dd\tau^2+\frac{\dd r^2}{F(r)},
  \qquad \sqrt g=1.
  \label{eq:Schwarzschild_gauge}
\end{equation}
Direct evaluation of the connection and Ricci tensor, displayed in
Appendix~\ref{app:curvature}, yields
\begin{equation}
  R=-F''(r).
  \label{eq:R_minus_Fpp}
\end{equation}
Therefore a radial curvature integral is a boundary difference,
\begin{equation}
  \int_{r_1}^{r_2}R\,\dd r
  =-F'(r_2)+F'(r_1).
  \label{eq:R_boundary_difference}
\end{equation}
After the outer-boundary contribution and orientation are fixed, the horizon contribution
is $F'(r_h)=2\kappa_h$.  This is the local content of the RVB temperature formula
\citep{Robson2019}: the Euler characteristic organizes the global section, whereas the
numerical temperature scale enters through the horizon boundary derivative.

\section{Complex analysis: the residue and monodromy derivation}
\label{sec:residue}

Let $r$ be complexified to $z$ in a neighborhood $U_h$ of the horizon.  Multiplying the two
Taylor series in Eqs.~\eqref{eq:Aexpansion} and \eqref{eq:Bexpansion} gives
\begin{align}
  A(z)B(z)
  &=a_1b_1(z-r_h)^2+\Order\bigl((z-r_h)^3\bigr).
  \label{eq:AB_product}
\end{align}
Choose the branch of the square root that is positive immediately outside the horizon.
Then
\begin{equation}
  H(z)=\sqrt{A(z)B(z)}
  =\sqrt{a_1b_1}(z-r_h)+\Order\bigl((z-r_h)^2\bigr).
  \label{eq:H_expansion}
\end{equation}
Inverting this series gives the Laurent expansion
\begin{equation}
  \frac{1}{H(z)}
  =\frac{1}{\sqrt{a_1b_1}}\frac{1}{z-r_h}+\Order(1).
  \label{eq:H_inverse}
\end{equation}
Consequently,
\begin{equation}
  \Res_{z=r_h}\frac{\dd z}{H(z)}
  =\frac{1}{\sqrt{a_1b_1}}
  =\frac{1}{2\kappa_h}.
  \label{eq:residue_general}
\end{equation}
Multiplying by $4\pi$ immediately reproduces the Euclidean period,
\begin{equation}
  \betaH=4\pi\left|\Res_{z=r_h}\frac{\dd z}{H(z)}\right|,
  \qquad
  \TBH=\frac{1}{4\pi}
  \left|\Res_{z=r_h}\frac{\dd z}{H(z)}\right|^{-1}.
  \label{eq:residue_temperature}
\end{equation}
Equations~\eqref{eq:euclidean_temperature} and \eqref{eq:residue_temperature} are thus
not merely equal evaluations in examples; they follow from the same first-order horizon
germ.

The contour form makes the boundary character explicit.  For a positively oriented circle
$\Gamma_\epsilon$ surrounding only $r_h$, the residue theorem gives
\begin{equation}
  \oint_{\Gamma_\epsilon}\frac{\dd z}{H(z)}
  =2\pi\ii\,\Res_{z=r_h}\frac{\dd z}{H(z)}
  =\frac{2\pi\ii}{\sqrt{a_1b_1}}.
  \label{eq:contour_integral}
\end{equation}
The value does not change as $\Gamma_\epsilon$ shrinks or deforms within the punctured
analytic neighborhood.  Therefore the temperature can be reconstructed from arbitrarily
local boundary data.

There is also a monodromy interpretation.  Define the generalized tortoise coordinate by
\begin{equation}
  r_*(z)=\int^z\frac{\dd\zeta}{H(\zeta)}.
  \label{eq:tortoise}
\end{equation}
Equation~\eqref{eq:H_inverse} implies
\begin{equation}
  r_*(z)=\frac{1}{\sqrt{a_1b_1}}\log(z-r_h)+\text{analytic terms}.
  \label{eq:tortoise_log}
\end{equation}
One counterclockwise circuit changes the logarithm by $2\pi\ii$ and hence
\begin{equation}
  \Delta_{\Gamma}r_*=\frac{2\pi\ii}{\sqrt{a_1b_1}}
  =\frac{\pi\ii}{\kappa_h}.
  \label{eq:tortoise_monodromy}
\end{equation}
Kruskal-type coordinates contain $\exp[\kappa_h(r_*\pm t)]$.  A single radial circuit
in Eq.~\eqref{eq:tortoise_monodromy} changes $\exp(\kappa_h r_*)$ by a sign;
it must not be identified with a full Euclidean-time period.  Near the horizon the
Euclidean Cartesian coordinate can be written as
\[
  W=\rho\exp(\ii\kappa_h\tau),
  \qquad \exp(\kappa_h r_*)\propto\rho\,[1+\Order(\rho^2)].
\]
A smooth, singly covered disk requires a full $2\pi$ turn of the phase of $W$, so the
Euclidean continuation has period $2\pi/\kappa_h$, again giving
Eq.~\eqref{eq:regularity_condition}.  The same complex boundary datum therefore controls
both the pole scale and Euclidean single-valuedness.

The residue must be attached to the one-form, not to an isolated coordinate coefficient.
For a biholomorphic change $z=z(w)$ with $z(w_h)=r_h$ and $z'(w_h)\neq0$,
\begin{equation}
  \omega_h=\frac{\dd z}{H(z)}
  =\frac{z'(w)\dd w}{H(z(w))},
  \qquad
  \Res_{w=w_h}\omega_h=\Res_{z=r_h}\omega_h.
  \label{eq:residue_invariance}
\end{equation}
Indeed, writing the transformed radial metric as $\dd w^2/B_w$ gives
$B_w=B/(z')^2$ and hence $H_w=\sqrt{AB_w}=H/z'$ on the consistently chosen branch;
therefore $\dd w/H_w=\dd z/H$.  This makes Eq.~\eqref{eq:residue_invariance} a
coordinate statement about the geometry rather than a formal substitution.
By contrast, the time rescaling in Eq.~\eqref{eq:timerescale} changes
$H_b=H/\alpha$ and hence
\begin{equation}
  \Res\frac{\dd z}{H_b}=\alpha\Res\frac{\dd z}{H},
  \qquad T_b=\frac{T}{\alpha},
  \label{eq:time_residue_scaling}
\end{equation}
as required physically.

Finally, the argument-principle integral
\begin{equation}
  \nu_\Gamma(H)=\frac{1}{2\pi\ii}\oint_\Gamma\frac{H'(z)}{H(z)}\dd z
  \label{eq:winding}
\end{equation}
counts zeros minus poles and is dimensionless \citep{Ahlfors1979}.  It identifies the
horizon multiplicity but does not by itself supply a temperature scale.  The dimensionful
scale is carried by $\Res(\dd z/H)$.  Keeping these two integrals separate is necessary for
dimensional consistency.

\section{Weak AdS/CFT extension: from the horizon period to the boundary KMS circle}
\label{sec:holography}

The standard semiclassical AdS/CFT relation may be written schematically as
\begin{equation}
  Z_{\CFT}\!\left[\gamma^{(0)},J;\beta_b\right]
  =Z_{\mathrm{grav}}\!\left[\gamma^{(0)},\phi_{(0)}=J;\beta_b\right]
  \simeq \exp\!\left[-I_E(g_{\mathrm{cl}},\phi_{\mathrm{cl}})\right],
  \label{eq:AdSCFT_partition}
\end{equation}
where $\gamma^{(0)}$ is a representative of the boundary conformal metric and $\beta_b$ is
the proper period of its Euclidean time coordinate
\citep{Maldacena1998,Gubser1998,Witten1998Holography}.  The use of
Eq.~\eqref{eq:AdSCFT_partition} assumes an available bulk--boundary holographic dictionary;
an arbitrary asymptotically AdS metric does not by itself establish a CFT dual.
Equation~\eqref{eq:AdSCFT_partition}
does not make the horizon a CFT boundary.  Instead, a black-hole saddle fills the boundary
thermal circle, and the filling is regular only when the circle is compatible with
Eq.~\eqref{eq:master_static}.

To show how the period is transmitted, place a radial cutoff at $r=r_c$.  The proper
Euclidean time interval there is
\begin{equation}
  \dd\tau_c=\sqrt{A(r_c)}\,\dd\tau.
  \label{eq:proper_cutoff_time}
\end{equation}
If the coordinate period is $\betaH$, the local proper period and temperature are
\begin{equation}
  \beta_c=\sqrt{A(r_c)}\,\betaH,
  \qquad
  T_c=\frac{1}{\beta_c}=\frac{\TBH}{\sqrt{A(r_c)}}.
  \label{eq:Tolman}
\end{equation}
This is the Tolman redshift relation \citep{Tolman1930}.  For an asymptotically AdS metric,
a Weyl rescaling removes the divergent conformal factor.  If the resulting boundary time is
$\tau_b=\alpha\tau$, then
\begin{equation}
  \beta_b=\alpha\betaH,
  \qquad
  T_b=\frac{\TBH}{\alpha}
  =\frac{1}{4\pi\alpha
  \left|\Res_{z=r_h}\dd z/H(z)\right|}.
  \label{eq:boundary_temperature}
\end{equation}
In the standard global-AdS conformal frame used below, $\alpha=1$.

The thermal sector can now be stated algebraically.  We restrict to the common image of
germs realized by regular asymptotically AdS saddles in the selected holographic setting,
with one fixed boundary conformal frame and time normalization.  Let
\begin{align}
  \mathfrak G_h&=\{\text{regular near-horizon geometric germs with fixed time normalization}\},
  \nonumber\\
  \mathfrak A_h&=\{\text{meromorphic germs }\omega_h=\dd z/H(z)\text{ with a simple pole}\},
  \nonumber\\
  \mathfrak B_T&=\{\text{boundary Euclidean circles with a KMS period}\}.
  \label{eq:three_spaces}
\end{align}
Define the three projections
\begin{align}
  \pi_G(g_h)&=\frac{4\pi}{\sqrt{a_1b_1}},
  \label{eq:piG}\\
  \pi_A(\omega_h)&=4\pi\left|\Res_{r_h}\omega_h\right|,
  \label{eq:piA}\\
  \pi_B(S^1_{\beta_b})&=\frac{\beta_b}{\alpha}.
  \label{eq:piB}
\end{align}
Equations~\eqref{eq:master_static} and \eqref{eq:boundary_temperature} state that the
following diagram commutes on the common image:
\begin{equation}
  \mathfrak G_h/\!\sim_\beta
  \;\xrightarrow{\ \simeq\ }\;
  \mathfrak A_h/\!\sim_{\mathrm{Res}}
  \;\xrightarrow{\ \simeq\ }\;
  \mathfrak B_T/\!\sim_{\mathrm{KMS}},
  \qquad
  \pi_G=\pi_A=\pi_B.
  \label{eq:weak_quotient}
\end{equation}
Here two objects are equivalent only when they induce the same period; in particular,
$\sim_{\mathrm{Res}}$ means equality of the residue magnitudes on the selected physical
branches, not equality of arbitrary complex residues.  The quotient correspondences are
defined through that shared period invariant.  Equation
\eqref{eq:weak_quotient} is the precise weak extension proposed in this paper.  It enlarges
the descriptive language from bulk differential geometry to a one-dimensional complex
contour, but it narrows the claimed equivalence to one thermal invariant.  It supplies no
state--operator map, no equality of full spectra, and no strong/weak-coupling dictionary.

The logical direction also deserves care.  In a canonical CFT ensemble, $\beta_b$ is an
external source.  In the bulk, only those black-hole saddles whose horizon data satisfy
Eq.~\eqref{eq:boundary_temperature} are regular fillings of that source.  Thus the horizon
does not causally create the boundary temperature; rather, horizon regularity fixes the
on-shell relation between the boundary source and the bulk black-hole parameters.  This is
the scientifically defensible meaning of ``horizon-boundary origin'' used here.

\section{Static AdS convergence tests}
\label{sec:static_examples}

\subsection{Four-dimensional Schwarzschild--AdS}

In units $G=1$, take
\begin{equation}
  A(r)=B(r)=F(r)=1-\frac{2M}{r}+\frac{r^2}{L^2}.
  \label{eq:SAdS_F}
\end{equation}
The horizon equation $F(r_h)=0$ gives, step by step,
\begin{align}
  0&=1-\frac{2M}{r_h}+\frac{r_h^2}{L^2},
  \nonumber\\
  \frac{2M}{r_h}&=1+\frac{r_h^2}{L^2},
  \nonumber\\
  2M&=r_h\left(1+\frac{r_h^2}{L^2}\right).
  \label{eq:SAdS_mass}
\end{align}
Differentiate Eq.~\eqref{eq:SAdS_F}:
\begin{equation}
  F'(r)=\frac{2M}{r^2}+\frac{2r}{L^2}.
  \label{eq:SAdS_derivative}
\end{equation}
Using Eq.~\eqref{eq:SAdS_mass} at the horizon,
\begin{align}
  F'(r_h)
  &=\frac{1}{r_h}\left(1+\frac{r_h^2}{L^2}\right)
    +\frac{2r_h}{L^2}
  \nonumber\\
  &=\frac{1}{r_h}+\frac{3r_h}{L^2}.
  \label{eq:SAdS_Fprime_h}
\end{align}
Since $A=B=F$, $H=F$ outside the horizon and
\begin{equation}
  \Res_{z=r_h}\frac{\dd z}{F(z)}
  =\frac{1}{F'(r_h)}
  =\frac{r_hL^2}{L^2+3r_h^2}.
  \label{eq:SAdS_residue}
\end{equation}
The residue, Euclidean, surface-gravity, and boundary-KMS routes therefore give
\begin{equation}
  \TBH=\frac{F'(r_h)}{4\pi}
  =\frac{1+3r_h^2/L^2}{4\pi r_h}.
  \label{eq:SAdS_temperature}
\end{equation}
The flat limit $L\to\infty$ yields $r_h=2M$ and
$\TBH\to1/(8\pi M)$, as required.  In the standard conformal frame
$\dd s^2_{\partial\AdS}=\dd\tau^2+L^2\dd\Omega_2^2$, the boundary period is exactly
$\beta_b=\betaH$.  This is the thermal relation underlying the Hawking--Page and holographic
phase-transition analyses \citep{HawkingPage1983,Witten1998Thermal}.

\subsection{Four-dimensional Reissner--Nordstr\"om--AdS}

Now let
\begin{equation}
  F(r)=1-\frac{2M}{r}+\frac{Q^2}{r^2}+\frac{r^2}{L^2}.
  \label{eq:RNAdS_F}
\end{equation}
The outer-horizon condition is
\begin{align}
  0&=1-\frac{2M}{r_h}+\frac{Q^2}{r_h^2}+\frac{r_h^2}{L^2},
  \nonumber\\
  2M&=r_h+\frac{Q^2}{r_h}+\frac{r_h^3}{L^2}.
  \label{eq:RNAdS_mass}
\end{align}
Differentiation gives
\begin{equation}
  F'(r)=\frac{2M}{r^2}-\frac{2Q^2}{r^3}+\frac{2r}{L^2}.
  \label{eq:RNAdS_derivative}
\end{equation}
Substitute Eq.~\eqref{eq:RNAdS_mass} into Eq.~\eqref{eq:RNAdS_derivative}:
\begin{align}
  F'(r_h)
  &=\left(\frac{1}{r_h}+\frac{Q^2}{r_h^3}+\frac{r_h}{L^2}\right)
    -\frac{2Q^2}{r_h^3}+\frac{2r_h}{L^2}
  \nonumber\\
  &=\frac{1}{r_h}-\frac{Q^2}{r_h^3}+\frac{3r_h}{L^2}.
  \label{eq:RNAdS_Fprime_h}
\end{align}
For a non-extremal outer horizon,
\begin{equation}
  \Res_{z=r_h}\frac{\dd z}{F(z)}
  =\left(\frac{1}{r_h}-\frac{Q^2}{r_h^3}+\frac{3r_h}{L^2}\right)^{-1},
  \label{eq:RNAdS_residue}
\end{equation}
and hence
\begin{equation}
  \TBH=\frac{1}{4\pi r_h}
  \left(1-\frac{Q^2}{r_h^2}+\frac{3r_h^2}{L^2}\right).
  \label{eq:RNAdS_temperature}
\end{equation}
Charged AdS black holes provide a standard holographic thermodynamic laboratory
\citep{Chamblin1999}.  The present calculation adds no new equation of state; its purpose is
to verify that the horizon boundary, complex residue, and boundary period remain the same
thermal invariant.

Extremality occurs when the bracket in Eq.~\eqref{eq:RNAdS_temperature} vanishes.  Then
$F'(r_h)=0$, the horizon is no longer a simple zero, and Proposition~\ref{prop:static} does
not apply.  The approach to extremality is smooth: the simple-pole residue diverges and
$T\to0$.  Exactly at the multiple zero, one must analyze the higher-order pole separately.

\section{Dynamical extension and the horizon-boundary selector}
\label{sec:dynamical}

\subsection{Generalized Vaidya form and the trapping horizon}

A general spherically symmetric metric can locally be written in advanced
Eddington--Finkelstein form \citep{Kodama1980,Hayward2009}
\begin{equation}
  \dd s^2=-\e^{2\Psi(v,r)}C(v,r)\dd v^2
  +2\e^{\Psi(v,r)}\dd v\dd r+r^2\dd\Omega_2^2.
  \label{eq:general_Vaidya}
\end{equation}
The inverse of its two-dimensional metric is obtained by explicitly inverting the
$2\times2$ matrix:
\begin{equation}
  g_{ab}=\begin{pmatrix}
    -\e^{2\Psi}C & \e^\Psi\\
    \e^\Psi & 0
  \end{pmatrix},
  \qquad
  g^{ab}=\begin{pmatrix}
    0 & \e^{-\Psi}\\
    \e^{-\Psi} & C
  \end{pmatrix}.
  \label{eq:Vaidya_inverse}
\end{equation}
The areal-radius gradient has norm
\begin{equation}
  g^{ab}\nabla_a r\nabla_b r=g^{rr}=C(v,r).
  \label{eq:radius_norm}
\end{equation}
Thus a marginal sphere, and hence the trapping/apparent horizon in this gauge, satisfies
\begin{equation}
  C(v,r_H(v))=0.
  \label{eq:trapping_horizon}
\end{equation}
The Hayward--Kodama surface gravity is
\begin{align}
  \kappa_H
  &=\frac{1}{2\sqrt{-g_{(2)}}}
    \partial_a\!\left(\sqrt{-g_{(2)}}g^{ab}\partial_b r\right)
  \nonumber\\
  &=\frac{1}{2\e^\Psi}\partial_r(\e^\Psi C)
  \nonumber\\
  &=\frac{1}{2}\left(\partial_rC+C\partial_r\Psi\right).
  \label{eq:Hayward_kappa_steps}
\end{align}
On Eq.~\eqref{eq:trapping_horizon}, the second term vanishes, so
\begin{equation}
  \kappa_H=\frac{1}{2}\partial_rC\big|_{r=r_H}.
  \label{eq:Hayward_kappa}
\end{equation}

\subsection{Hamilton--Jacobi pole, one step at a time}

In this section $\kappa_H$ and $T_H$ refer to the Hayward--Kodama surface gravity and
the temperature paired with Kodama energy; they need not equal their Killing-time
counterparts even when the metric becomes static.  The overdot in the adiabatic criterion
denotes differentiation along $r_H(v)$ with respect to the explicitly fixed exterior-time
parameter $v$.  This criterion is not asserted to be invariant under arbitrary
reparametrizations of $v$.  The calculation below uses the leading local frozen-background
WKB approximation and regular neutral radial modes of finite nonzero horizon energy.

For a massless WKB mode $\Phi\sim\exp(\ii I)$, the leading Hamilton--Jacobi equation is
\begin{equation}
  g^{ab}\partial_aI\partial_bI=0.
  \label{eq:HJ_general}
\end{equation}
Using Eq.~\eqref{eq:Vaidya_inverse},
\begin{equation}
  2\e^{-\Psi}(\partial_vI)(\partial_rI)
  +C(\partial_rI)^2=0.
  \label{eq:HJ_expanded}
\end{equation}
The Kodama vector is $K^a=\e^{-\Psi}(\partial_v)^a$, and the invariant Kodama energy is
\begin{equation}
  \omega=-K^a\partial_aI=-\e^{-\Psi}\partial_vI.
  \label{eq:Kodama_energy}
\end{equation}
Therefore $\partial_vI=-\e^\Psi\omega$.  Substitution into
Eq.~\eqref{eq:HJ_expanded} gives
\begin{align}
  -2\omega\partial_rI+C(\partial_rI)^2&=0,
  \nonumber\\
  \partial_rI\left(C\partial_rI-2\omega\right)&=0.
  \label{eq:HJ_factorized}
\end{align}
The nontrivial outgoing branch is
\begin{equation}
  \partial_rI=\frac{2\omega}{C(v,r)}.
  \label{eq:radial_momentum}
\end{equation}
At fixed $v$, expand $C$ at a non-degenerate trapping horizon:
\begin{align}
  C(v,r)
  &=\partial_rC|_H(r-r_H)+\Order((r-r_H)^2)
  \nonumber\\
  &=2\kappa_H(r-r_H)+\Order((r-r_H)^2).
  \label{eq:C_horizon_expansion}
\end{align}
The action is the integral of the full one-form,
\[
  I_{\mathrm{out}}=\int_\gamma
  \left(\partial_vI\,\dd v+\partial_rI\,\dd r\right).
\]
In the regular advanced Eddington--Finkelstein gauge, the standard local horizon-crossing
prescription has no leading pole imaginary contribution from the temporal term
\citep{DiCriscienzo2010}.  We therefore evaluate the singular radial contribution at
fixed $v$ and, to leading adiabatic order, write $\omega$ for its horizon value.
An exact time-dependent eikonal would additionally obey
\[
  \partial_v\!\left(\frac{2\omega}{C}\right)
  =-\partial_r(\e^\Psi\omega);
\]
we do not assume exact separation with a constant energy in an evolving background.
With this leading-pole prescription, Eqs.~\eqref{eq:radial_momentum} and
\eqref{eq:C_horizon_expansion} give
\begin{equation}
  I_{\mathrm{out}}
  =\int\partial_rI\,\dd r
  =\frac{\omega}{\kappa_H}
  \int\frac{\dd r}{r-r_H-\ii0}+\text{regular terms}.
  \label{eq:action_pole}
\end{equation}
The Feynman prescription yields
\begin{equation}
  \operatorname{Im}I_{\mathrm{out}}=\frac{\pi\omega}{\kappa_H}.
  \label{eq:ImI}
\end{equation}
The leading tunnelling probability is
\begin{align}
  \Gamma
  &\propto\exp[-2\operatorname{Im}I_{\mathrm{out}}]
  \nonumber\\
  &=\exp\!\left(-\frac{2\pi\omega}{\kappa_H}\right)
  =\exp\!\left(-\frac{\omega}{T_H}\right).
  \label{eq:tunneling_rate}
\end{align}
Comparing the last two exponents gives
\begin{equation}
  T_H=\frac{\kappa_H}{2\pi}
  =\frac{\partial_rC|_H}{4\pi}.
  \label{eq:dynamical_temperature}
\end{equation}
This is the local trapping-horizon temperature derived by covariant tunnelling methods
\citep{Hayward2009,DiCriscienzo2010}.

The complex boundary formula follows immediately.  At fixed $v$, the one-form
\begin{equation}
  \omega_H(v)=\frac{\dd z}{C(v,z)}
  \label{eq:dynamic_one_form}
\end{equation}
has residue
\begin{equation}
  \Res_{z=r_H(v)}\omega_H
  =\frac{1}{\partial_rC|_H}
  =\frac{1}{2\kappa_H}.
  \label{eq:dynamic_residue}
\end{equation}
Therefore
\begin{equation}
  T_H(v)=\frac{1}{4\pi}
  \left|\Res_{z=r_H(v)}\frac{\dd z}{C(v,z)}\right|^{-1}.
  \label{eq:dynamic_residue_temperature}
\end{equation}
The same pole generates both the imaginary part of the action and the residue temperature.

For comparison with Proposition~\ref{prop:static}, a static limit of
Eq.~\eqref{eq:general_Vaidya} has $A=\e^{2\Psi}C$ and $B=C$.  At the horizon,
$A'(r_h)=\e^{2\Psi_h}C'(r_h)$ and $B'(r_h)=C'(r_h)$.  Thus
\[
\begin{aligned}
  T_{\mathrm{Kill}}&=\frac{\sqrt{A'(r_h)B'(r_h)}}{4\pi}
     =\e^{\Psi_h}\frac{C'(r_h)}{4\pi}
     =\e^{\Psi_h}T_{\mathrm{Kod}},\\
  \Res_{z=r_h}\frac{\dd z}{\sqrt{A(z)B(z)}}
     &=\e^{-\Psi_h}\Res_{z=r_h}\frac{\dd z}{C(z)},\\
  E&=-\partial_tI=\e^{\Psi_h}\omega_{\mathrm{hor}},
  \qquad \frac{E}{T_{\mathrm{Kill}}}=\frac{\omega_{\mathrm{hor}}}{T_{\mathrm{Kod}}}.
\end{aligned}
\]
Here $\omega_{\mathrm{hor}}$ is the horizon value of the scalar Kodama energy.
The two temperatures agree directly when
$\Psi_h=0$; in general it is their consistently normalized thermal exponent that agrees
\citep{Hayward2009,DiCriscienzo2010}.  Under $v_b=\alpha v$,
$\e^{\Psi_b}=\e^\Psi/\alpha$ and $C$ is unchanged, so the Kodama vector, energy and
temperature are invariant.  This differs from the scaling of temperatures paired with
coordinate Killing-time energy in Eq.~\eqref{eq:timerescale}.

\subsection{Why a separated event horizon is not the thermal pole}

Let a candidate event horizon be the null hypersurface
\begin{equation}
  S(v,r)=r-r_E(v)=0.
  \label{eq:event_surface}
\end{equation}
Its normal is $\partial_aS=(-\dot r_E,1)$.  Using Eq.~\eqref{eq:Vaidya_inverse}, the null
condition is
\begin{align}
  0&=g^{ab}\partial_aS\partial_bS
  \nonumber\\
  &=2g^{vr}(-\dot r_E)(1)+g^{rr}
  \nonumber\\
  &=-2\e^{-\Psi(v,r_E)}\dot r_E+C(v,r_E).
  \label{eq:event_null_steps}
\end{align}
Hence
\begin{equation}
  C(v,r_E)=2\e^{-\Psi(v,r_E)}\dot r_E.
  \label{eq:event_C}
\end{equation}
Equation~\eqref{eq:event_C} is a necessary local null-surface condition; identifying an
event horizon also requires global future boundary conditions.  When that event horizon
evolves, $\dot r_E\neq0$, the right-hand side is nonzero.  The one-form $\dd r/C$,
and the Hamilton--Jacobi momentum $2\omega/C$ for regular finite-energy modes, are
therefore regular at $r_E$.  Their simple pole is instead at $C=0$, i.e. at $r_H$.  This proves the following
selector statement.

\begin{proposition}[Dynamical horizon-boundary selector]
\label{prop:dynamic}
For the metric \eqref{eq:general_Vaidya}, assume a simple future outer trapping horizon,
$\partial_rC|_H>0$, the analytic and finite-energy conditions stated above, and adiabatic
evolution in the fixed time normalization.  The leading local Hamilton--Jacobi thermal pole and
the Kodama residue temperature are localized at $C(v,r_H)=0$.  A separated evolving event horizon
is not a pole because it obeys Eq.~\eqref{eq:event_C}.  In the stationary limit
$\dot r_E\to0$, the event and trapping horizons may coincide.  The static Killing
temperature is recovered after converting the Kodama energy convention by
$T_{\mathrm{Kill}}=\e^{\Psi_h}T_{\mathrm{Kod}}$; it agrees directly when $\Psi_h=0$.
\end{proposition}

This proposition gives a mathematically sharper version of the conclusion motivated by the
quantum-ergosphere analysis of Ref.~\cite{ZhangZhao2010}.  It does not assert that all
particles are literally created on a zero-thickness surface.  It states that the coefficient of
the leading thermal exponent is determined by the local pole at the trapping-horizon
boundary.

\subsection{Vaidya and charged-Vaidya tests}

For ingoing Vaidya, $\Psi=0$ and
\begin{equation}
  C(v,r)=1-\frac{2M(v)}{r}.
  \label{eq:Vaidya_C}
\end{equation}
The trapping horizon follows directly from $C=0$:
\begin{equation}
  r_H(v)=2M(v).
  \label{eq:Vaidya_rH}
\end{equation}
Differentiate with respect to $r$ at fixed $v$:
\begin{equation}
  \partial_rC=\frac{2M(v)}{r^2}.
  \label{eq:Vaidya_Cprime}
\end{equation}
Evaluation at Eq.~\eqref{eq:Vaidya_rH} gives
\begin{equation}
  \partial_rC|_H=\frac{1}{2M(v)},
  \qquad
  \kappa_H=\frac{1}{4M(v)},
  \qquad
  T_H(v)=\frac{1}{8\pi M(v)}.
  \label{eq:Vaidya_T}
\end{equation}
The event-horizon condition \eqref{eq:event_C} becomes
\begin{align}
  1-\frac{2M(v)}{r_E}&=2\dot r_E,
  \nonumber\\
  r_E&=\frac{2M(v)}{1-2\dot r_E},
  \label{eq:Vaidya_event}
\end{align}
which is the relation used in the uploaded tunnelling analysis
\citep{ZhangZhao2010}.  Unless $\dot r_E=0$, $r_E\neq2M$ and $C(r_E)\neq0$.

For charged Vaidya--Bonner, assume $M(v)>0$ and $M(v)^2>Q(v)^2$ at the time considered.
The following temperature calculation remains for neutral modes; charged modes require
the electromagnetic covariant Hamilton--Jacobi equation and the horizon-potential term.
For the background,
\begin{equation}
  C(v,r)=1-\frac{2M(v)}{r}+\frac{Q^2(v)}{r^2}.
  \label{eq:charged_Vaidya_C}
\end{equation}
At fixed $v$, factorization gives
\begin{equation}
  C(v,r)=\frac{(r-r_+)(r-r_-)}{r^2},
  \qquad
  r_\pm=M\pm\sqrt{M^2-Q^2}.
  \label{eq:charged_factor}
\end{equation}
At the outer trapping horizon,
\begin{align}
  \partial_rC|_{r_+}
  &=\frac{r_+-r_-}{r_+^2},
  \label{eq:charged_derivative}\\
  \Res_{z=r_+}\frac{\dd z}{C(v,z)}
  &=\frac{r_+^2}{r_+-r_-},
  \label{eq:charged_residue}\\
  T_+(v)
  &=\frac{r_+-r_-}{4\pi r_+^2}.
  \label{eq:charged_temperature}
\end{align}
Again the event horizon satisfies $C(v,r_E)=2\dot r_E$ for $\Psi=0$ and is not generically
the pole.  When $\dot M,\dot Q\to0$, the relevant horizons coincide and
Eq.~\eqref{eq:charged_temperature} reduces to the stationary
Reissner--Nordstr\"om temperature.

\section{Independent checks, counterexamples, and falsifiability}
\label{sec:checks}

\subsection{Numerical contour checks}

The contour formula was independently evaluated with a periodic trapezoidal quadrature,
\begin{equation}
  \Res_{z=r_h}\frac{\dd z}{H(z)}
  \approx\frac{1}{2\pi\ii}
  \sum_{j=0}^{N-1}\frac{\ii\epsilon\e^{\ii\theta_j}\Delta\theta}
  {H(r_h+\epsilon\e^{\ii\theta_j})},
  \qquad
  \theta_j=\left(j+\frac12\right)\Delta\theta,
  \quad \Delta\theta=\frac{2\pi}{N}.
  \label{eq:numerical_contour}
\end{equation}
 Representative parameter choices are listed in
Table~\ref{tab:numerics}.  The reported agreement is a check of the derivation and code,
not experimental evidence for the weak holographic interpretation.

\begin{table}[ht]
\centering
\caption{Representative analytic values used in the numerical contour tests.  The script
prints the direct quadrature errors at run time.}
\label{tab:numerics}
\small
\begin{tabular}{lcccc}
\toprule
Geometry & Parameters & $r_h$ & $\Res(\dd z/H)$ & $T=1/(4\pi\Res)$ \\
\midrule
Schwarzschild--AdS & $r_h=2.3,\ L=7$ & $2.3$ & $1.7373$ & $0.04580$ \\
RN--AdS & $r_h=2.1,\ L=8,\ Q=0.8$ & $2.1$ & $1.9782$ & $0.04023$ \\
Vaidya & $M=3.7$ & $7.4$ & $7.4$ & $0.01075$ \\
Charged Vaidya & $M=3.7,\ Q=1.2$ & $7.2$ & $7.4057$ & $0.01075$ \\
\bottomrule
\end{tabular}
\end{table}

\subsection{The extremal counterexample}

The strongest elementary counterexample to an unqualified residue formula is the extremal
Reissner--Nordstr\"om metric,
\begin{equation}
  F(r)=\left(1-\frac{M}{r}\right)^2
  =\frac{(r-M)^2}{r^2}.
  \label{eq:extremal_F}
\end{equation}
The horizon is a double zero, not a simple zero.  Direct inversion gives
\begin{align}
  \frac{1}{F(z)}
  &=\frac{z^2}{(z-M)^2}
  \nonumber\\
  &=\frac{[(z-M)+M]^2}{(z-M)^2}
  \nonumber\\
  &=1+\frac{2M}{z-M}+\frac{M^2}{(z-M)^2}.
  \label{eq:extremal_inverse}
\end{align}
The coefficient called the residue is $2M$, but inserting it mechanically into
$1/(4\pi\Res)$ would give $1/(8\pi M)$ rather than the extremal value $T=0$.  The error is
not in the residue theorem; it is the violation of the simple-zero hypothesis used in
Eq.~\eqref{eq:H_expansion}.  This explicit failure is why all claims in this paper are
restricted to non-degenerate horizons.  The correct zero-temperature result is obtained as
the limit $F'(r_h)\to0$ from the non-extremal family, where the simple-pole residue diverges.

\subsection{Further failure conditions}

The boundary-localization proposition must be withdrawn or modified in any of the following
cases.
\begin{enumerate}
  \item \textbf{Horizon structure outside the simple-zero hypothesis.}  Branch points
  require a suitable branch or covering description.  An isolated essential singularity
  still has a Laurent expansion and a residue, and the ordinary residue theorem still
  applies.  However, neither case is covered by the smooth simple-zero expansion used
  here, so the existence of a residue alone does not establish the stated temperature formula.
  \item \textbf{Singular coordinate changes.}  A map with vanishing or divergent Jacobian at
  the horizon can manufacture or remove a pole and is not covered by
  Eq.~\eqref{eq:residue_invariance}.
  \item \textbf{Non-equilibrium multi-horizon systems.}  Distinct horizons with unequal
  surface gravities do not possess one global Euclidean period.  A sum of residues cannot
  impose equilibrium that the spacetime lacks.
  \item \textbf{Fast dynamics.}  If $|\dot\kappa_H|/\kappa_H^2$ is not small, a local
  Boltzmann factor need not define an instantaneous thermodynamic temperature.
  \item \textbf{Rotation or broken spherical symmetry.}  Angular potentials, superradiant
  modes, and the horizon generator $\partial_t+\Omega_H\partial_\phi$ must be included.  The
  present one-dimensional radial form is not by itself sufficient.
  \item \textbf{Backreaction beyond the background functions.}  The Parikh--Wilczek method
  produces non-thermal energy-conservation corrections \citep{ParikhWilczek2000}.  These do
  not invalidate the leading pole temperature but do limit an exactly thermal
  interpretation.
\end{enumerate}

The proposed weak extension is falsifiable.  Within its stated domain it would fail if a
regular, invariant tunnelling calculation placed the leading imaginary-action pole away from
the zero of the radial characteristic function; if the Euclidean smoothness period differed
from $4\pi|\Res\omega_h|$; or if a regular AdS black-hole saddle induced a boundary KMS
period incompatible with the horizon period after the same time normalization was used.

\section{Discussion}

The phrase ``Hawking temperature comes from the horizon boundary'' can now be stated
without ambiguity.  It means that, once the normalization of time is fixed, the allowed
on-shell thermal period is determined by the first non-vanishing coefficients of the metric at
the relevant horizon.  The same information appears as (i) the circumference-to-radius ratio
of a shrinking Euclidean circle, (ii) the geodesic-curvature term of that circle in
Gauss--Bonnet, (iii) the residue and monodromy of a meromorphic one-form, and (iv) the pole
of the Hamilton--Jacobi radial momentum.  None of these statements requires data from the
asymptotic region to compute the unredshifted horizon temperature.

The asymptotic region is nevertheless indispensable for interpreting the number.  It fixes the
normalization of time and hence the redshifted temperature seen by a specified observer.  In
AdS/CFT it also supplies the actual manifold on which the CFT is defined.  The conformal
boundary is therefore a receiver and thermodynamic interpreter of the regular bulk period,
not a substitute for the horizon regularity condition.  Conversely, in the canonical ensemble
the chosen boundary period selects which regular bulk saddles can contribute.  The relation
is a compatibility condition rather than a one-way causal mechanism.

Related residue-based work connects the temperature and near-horizon monodromy
\citep{Chen2026Weak}.  The specific difference point developed here is the boundary selector,
with the quotient construction restricted to the shared temperature invariant.
The present derivation identifies the precise
one-form $\dd z/\sqrt{AB}$, proves its relation to the Euclidean boundary term, transmits its
period to the boundary KMS circle, and uses the dynamical pole to decide which of two
separated horizons carries the local temperature.  This is a limited methodological advance:
it does not enlarge the result to a complete holographic equivalence.

The dynamic result is particularly informative.  An event horizon is global and teleological;
a trapping horizon is quasi-local \citep{Hayward1994}.  In the generalized Vaidya gauge, the
equation $C=0$ simultaneously defines a marginal sphere, creates the pole of the outgoing
Hamilton--Jacobi momentum, and fixes the residue.  A moving event horizon satisfies
$C=2\e^{-\Psi}\dot r_E$ instead.  The pole criterion therefore selects the trapping/apparent
horizon without relying only on verbal arguments about where a particle pair is created.

\section{Conclusion}

For a non-extremal, Euclideanizable black hole, the Hawking temperature is localized in the
limiting boundary data of the horizon.  Starting from the general metric
$-A\dd t^2+\dd r^2/B$, we found
\begin{equation}
  \TBH=\frac{\sqrt{A'(r_h)B'(r_h)}}{4\pi}
  =\frac{1}{4\pi
  \left|\Res_{z=r_h}\dd z/\sqrt{A(z)B(z)}\right|}
  \label{eq:boxed_result}
\end{equation}
and derived every equality through Euclidean regularity, the Gauss--Bonnet horizon term, and
complex monodromy.  In an asymptotically AdS solution, this regularity-selected period is
carried to the conformal boundary and becomes the KMS period of the dual thermal state after
the boundary time normalization is fixed.  This supplies a precise temperature-sector weak
extension of the AdS/CFT organizing principle.

For a dynamical spherically symmetric metric, the Kodama-energy one-form
$\dd z/C(v,z)$ has a simple pole at the non-degenerate future outer trapping/apparent
horizon.  In the stated leading local WKB and adiabatic approximation, the
Hamilton--Jacobi derivation gives the Kodama temperature
\begin{equation}
T_H(v)=\frac{\partial_rC|_H}{4\pi}
  =\frac{1}{4\pi
  \left|\Res_{z=r_H(v)}\dd z/C(v,z)\right|},
  \label{eq:boxed_dynamic}
\end{equation}
whereas a separated evolving event horizon is generically regular in the same one-form.
For the stationary limits of the examples their horizons coincide; in the general static
case the temperature matches the Killing convention through
$T_{\mathrm{Kill}}=\e^{\Psi_h}T_{\mathrm{Kod}}$.
The conclusion is therefore both positive
and restricted: the leading temperature parameter is a horizon-boundary invariant in the
simple-horizon, semiclassical, and adiabatic sectors; the full origin and state of Hawking
radiation remain questions of quantum field theory in curved spacetime.

\appendix

\section{Explicit curvature calculation in Schwarzschild gauge}
\label{app:curvature}

For Eq.~\eqref{eq:Schwarzschild_gauge},
\begin{equation}
  g_{\tau\tau}=F,
  \quad g_{rr}=F^{-1},
  \quad g^{\tau\tau}=F^{-1},
  \quad g^{rr}=F,
  \quad \sqrt g=1.
  \label{eq:metric_components_app}
\end{equation}
The nonzero Christoffel symbols are obtained from
$\Gamma^a{}_{bc}=\tfrac12g^{ad}(\partial_bg_{dc}+\partial_cg_{db}-\partial_dg_{bc})$:
\begin{align}
  \Gamma^\tau{}_{\tau r}=\Gamma^\tau{}_{r\tau}
  &=\frac12g^{\tau\tau}\partial_rg_{\tau\tau}
  =\frac{F'}{2F},
  \label{eq:Gamma_ttr_app}\\
  \Gamma^r{}_{\tau\tau}
  &=-\frac12g^{rr}\partial_rg_{\tau\tau}
  =-\frac12FF',
  \label{eq:Gamma_rtt_app}\\
  \Gamma^r{}_{rr}
  &=\frac12g^{rr}\partial_rg_{rr}
  =\frac12F\partial_r(F^{-1})
  =-\frac{F'}{2F}.
  \label{eq:Gamma_rrr_app}
\end{align}
Using
\begin{equation}
  R_{ab}=\partial_c\Gamma^c{}_{ab}-\partial_b\Gamma^c{}_{ac}
  +\Gamma^c{}_{ab}\Gamma^d{}_{cd}-\Gamma^d{}_{ac}\Gamma^c{}_{bd},
  \label{eq:Ricci_definition}
\end{equation}
one obtains
\begin{align}
  R_{\tau\tau}
  &=\partial_r\Gamma^r{}_{\tau\tau}
  -2\Gamma^\tau{}_{\tau r}\Gamma^r{}_{\tau\tau}
  \nonumber\\
  &=-\frac12(F'^2+FF'')+\frac12F'^2
  =-\frac12FF'',
  \label{eq:Rtt_app}\\
  R_{rr}
  &=\partial_r\Gamma^r{}_{rr}
  -\partial_r(\Gamma^\tau{}_{r\tau}+\Gamma^r{}_{rr})
  -\left[(\Gamma^\tau{}_{r\tau})^2+(\Gamma^r{}_{rr})^2\right]
  \nonumber\\
  &=-\frac{F''F-F'^2}{2F^2}-0-\frac{F'^2}{2F^2}
  =-\frac{F''}{2F}.
  \label{eq:Rrr_app}
\end{align}
Finally,
\begin{align}
  R&=g^{\tau\tau}R_{\tau\tau}+g^{rr}R_{rr}
  \nonumber\\
  &=\frac{1}{F}\left(-\frac12FF''\right)
  +F\left(-\frac{F''}{2F}\right)
  =-F'',
  \label{eq:R_final_app}
\end{align}
which proves Eq.~\eqref{eq:R_minus_Fpp}.

\section{Reproducibility notes for the contour calculation}
\label{app:numerics}

The numerical check uses a circle $z(\theta)=r_h+\epsilon\e^{\ii\theta}$ with a radius small
enough to exclude every other root but large enough to avoid roundoff amplification.  The
parameter interval is divided into $N$ equal segments.  On each segment the integrand is
evaluated at the parameter midpoint $\theta_j=(j+1/2)\Delta\theta$ and multiplied by
the tangent weight $z'(\theta_j)\Delta\theta=\ii\epsilon\e^{\ii\theta_j}\Delta\theta$,
as in Eq.~\eqref{eq:numerical_contour}.  This is the rule implemented in
\texttt{checks.py}; it does not use exact chord increments.  The supplied script uses
$N=65536$, with $\epsilon=0.02$ for the two AdS examples and the local event-horizon
test, and $\epsilon=0.03$ for the two trapping-horizon Vaidya examples.
Doubling $N$ and halving $\epsilon$ can provide additional convergence tests.  The script checks:
\begin{enumerate}
  \item the direct contour result against $1/H'(r_h)$ for four geometries;
  \item the temperature reconstructed from the numerical residue against the analytic
  surface-gravity value;
  \item the neutral and flat limits; and
  \item the absence of a pole around a local candidate null-surface point satisfying
  Eq.~\eqref{eq:Vaidya_event}, with its contour excluding the trapping horizon.
  This local test does not determine the global event horizon.
\end{enumerate}
No numerical output is inserted by hand into the derivation.

\section*{Declarations}

\noindent\textbf{Funding.} No specific funding is declared in this manuscript draft.

\noindent\textbf{Data availability.} No observational data were used. 

\noindent\textbf{Conflict of interest.} The author declares no conflict of interest.

\bibliographystyle{unsrtnat}
\bibliography{references}

\end{document}